\documentclass[9pt, conference]{IEEEtran}
\IEEEoverridecommandlockouts

\usepackage{cite}
\usepackage{amsmath,amssymb,amsfonts}
\usepackage{algorithmic}
\usepackage{graphicx}
\usepackage{textcomp}
\usepackage{hyperref}

\usepackage{multirow}
\usepackage[table]{xcolor}
\usepackage{booktabs}

\def\BibTeX{{\rm B\kern-.05em{\sc i\kern-.025em b}\kern-.08em
    T\kern-.1667em\lower.7ex\hbox{E}\kern-.125emX}}
\begin{document}

\title{STA-V2A: Video-to-Audio Generation with Semantic and Temporal Alignment}

\author{
    \IEEEauthorblockN{Yong Ren$^{1}$, Chenxing Li\IEEEauthorrefmark{1}\IEEEauthorrefmark{2}$^{1}$\thanks{*Corresponding author}, Manjie Xu$^{1,3}$, Wei Liang$^{3}$, Yu Gu$^{1}$, Rilin Chen$^{1}$, Dong Yu\IEEEauthorrefmark{2}$^{2}$}
    \IEEEauthorblockA{$^1$ Tencent AI Lab, Beijing, China}
    \IEEEauthorblockA{$^2$ Tencent AI Lab, Seattle, USA}
    \IEEEauthorblockA{$^3$ Beijing Institute of Technology}
    \IEEEauthorblockA{\IEEEauthorrefmark{2} lichenxing007@gmail.com, dongyu@ieee.org}
}

\maketitle

\begin{abstract}
Visual and auditory perception are two crucial ways humans experience the world. Text-to-video generation has made remarkable progress over the past year, but the absence of harmonious audio in generated video limits its broader applications. In this paper, we propose Semantic and Temporal Aligned Video-to-Audio (STA-V2A), an approach that enhances audio generation from videos by extracting both local temporal and global semantic video features and combining these refined video features with text as cross-modal guidance. To address the issue of information redundancy in videos, we propose an onset prediction pretext task for local temporal feature extraction and an attentive pooling module for global semantic feature extraction. To supplement the insufficient semantic information in videos, we propose a Latent Diffusion Model with Text-to-Audio priors initialization and cross-modal guidance. We also introduce Audio-Audio Align, a new metric to assess audio-temporal alignment. Subjective and objective metrics demonstrate that our method surpasses existing Video-to-Audio models in generating audio with better quality, semantic consistency, and temporal alignment. The ablation experiment validated the effectiveness of each module. Audio samples are available at https://y-ren16.github.io/STAV2A.
\end{abstract}


\begin{figure*}[htbp]
\centering
  \centerline{\includegraphics[width=0.85\linewidth]{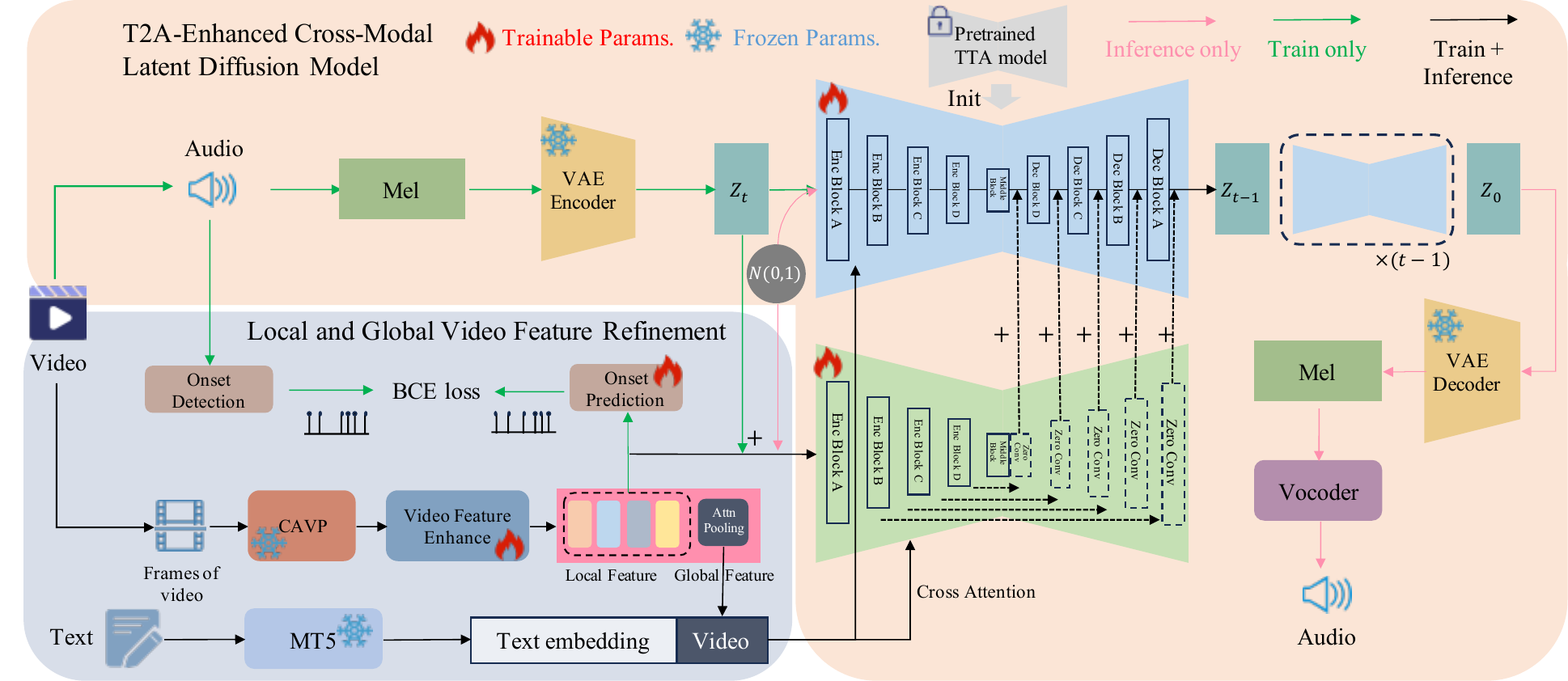}}
  \caption{
  Overview of the STA-V2A framework. The local and global video feature refinement module extracts local temporal and global semantic video features through onset prediction loss and an attention pooling module. 
  The pre-trained T2A model initializes the LDM, with text and global video features serving as semantic conditions introduced via cross-attention and local video features acting as temporal conditions introduced through an adapter.
  }
  \label{fig:framwork}
\end{figure*}

\section{Introduction}
Generating audio harmonizing with the video is an important task in generative artificial intelligence.
Currently, there are three main directions to tackle this challenge.
Text-To-Audio (T2A) \cite{liu2023audioldm, liu2024audioldm, ghosal2023text} generates audio conditioned on text. By leveraging the semantic information in the text descriptions of a video, these methods can generate high-quality audio with good semantic consistency. 
Text-To-Video-with-Audio (T2VA) \cite{ruan2023mm, xing2024seeing, mao2024tavgbench, hayakawa2024discriminator} simultaneously generates video and audio conditioned on text, which demonstrates temporal consistency. 
Video-To-Audio (V2A) \cite{iashin2021taming, sheffer2023hear, luo2024diff, xu2024video, pascual2024masked, wang2024frieren} generates audio conditioned on video features, which can effectively utilize the semantic and temporal information contained in videos, resulting in well-aligned audio.
Some recent approaches \cite{mo2024text, zhang2024foleycrafter} incorporate text conditions as a supplement to the video semantic information in V2A, enhancing the generated audio's semantic consistency.

Each of these three categories of methods faces challenges in generating high-quality audio semantically and temporally aligned with video.
T2A methods often struggle with temporal alignment due to a lack of video-related temporal information in the input. 
T2VA methods employ joint audio-visual cross-modal generation through alignment methods across different modalities, aligning audio and video elements in the latent space. However, the joint generation also leads to increased model complexity and a decline in generation quality.
V2A methods use video as a condition, effectively leveraging the semantic and temporal information within the video. 
Due to the large amount of information in videos, extracting video features is crucial for these methods. Diff-Foley \cite{luo2024diff} introduced a contrastive audio-visual pre-training (CAVP) method to learn video representations, achieving better temporal alignment. VTA-LDM \cite{xu2024video} compared various video feature extraction methods and attempted to generate semantically and temporally aligned audio based on video through end-to-end training. 
Due to the redundancy of various information in the video and the interference from some audio-irrelevant information, relying solely on single video features extracted from a pre-trained model does not effectively guide the semantic and temporal aligned audio generation.
The text can supplement the video by providing additional semantic information, which helps improve the semantic consistency of the generated audio.
Mo et al. were the first to use both text and video as conditions for audio generation \cite{mo2024text}. Due to insufficient utilization of text and video features, the quality of the generated audio is limited.
Concurrent work FoleyCrafter \cite{zhang2024foleycrafter} employs a semantic adapter and a temporal controller to achieve semantic and temporal alignment. However, an additional labelled audio-visual dataset is required during training, which may result in reduced performance when generalizing to universal audio generation.

To generate semantically and temporally aligned audio for the target video, we propose Semantic and Temporal Aligned Video-to-Audio (STA-V2A). 
To tackle the issue of interference from redundant information in video features, we extract local temporal and global semantic features of videos.
For the local feature, we propose an onset predict pretext task that predicts audio onset from video features \cite{comunita2024syncfusion,yariv2024diverse}. For the global feature, we propose a trainable attentive pooling module \cite{ali2022video} to extract the semantic feature of the video.
To address the issue of insufficient semantic information in video features and ensure the quality of generated audio, we employ prior knowledge of the pre-trained T2A model to initialize the diffusion model for generating high-quality audio. Additionally, we use both text and video features as cross-modal guidance to ensure temporal and semantic alignment.
The contributions are as follows:
\begin{itemize}
\item 
\textbf{Local and Global Video Feature Refinement}: 
This paper proposes an onset prediction pretext task to obtain local temporal features of video alongside a trainable attentive pooling module to acquire global semantic features of video, achieving refinement and detailing of temporal and semantic information in videos.
\item 
\textbf{T2A-Enhanced Cross-Modal Latent Diffusion Model}: This paper introduces a Latent Diffusion Model (LDM) framework where the initialization with T2A ensures high-quality audio generation and the cross-modal guidance between text and video ensures semantic and temporal consistency in generated videos.
\item 
\textbf{New Evaluation Metric for Audio Temporal Alignment}: This paper introduces a new metric, Audio-Audio Alignment (AA-Align), addressing the lack of effective metrics for the temporal alignment of audio.
\end{itemize}
Comprehensive experiments have fully demonstrated that STA-V2A surpasses existing V2A methods regarding generation quality, semantic consistency, and temporal alignment.

\section{Method}
\subsection{Overall Framework}\label{framwork}
As shown in Fig.~\ref{fig:framwork}, The mel-spectrogram is compressed into latent variables $z$ through the Variational Autoencoder (VAE) \cite{kingma2013auto} encoder. The LDM generates $z$, which is then processed by the VAE Decoder and HiFi-GAN vocoder \cite{kong2020hifi} to recover the audio. 
We propose a \textbf{local and global video feature refinement} method to obtain video representations rich in temporal and semantic information.
Then, we propose \textbf{LDM with T2A prior and cross-modal guidance}, leveraging T2A prior knowledge and cross-modal semantic and temporal information from text and video to generate audio. 
Lastly, to better evaluate the temporal consistency of generated audio, we propose a \textbf{new evaluation metric, AA-Align}. 
In the following sections, we will introduce these three innovations contributions.

\subsection{Local and Global Video Feature Refinement}\label{video_feature}
Videos contain a wealth of audio-related semantic and temporal features, making video feature extraction crucial for V2A tasks. In Diff-Foley \cite{luo2024diff}, the CAVP method learns temporally aligned information between audio and video through audio-visual contrastive pre-training. We employ the CAVP pre-trained model to extract initial video features $e_v$ and then design two approaches to separately obtain local temporal features $e_{lv}$ and global semantic features $e_{gv}$ of the video. These enhanced video features are subsequently used as inputs for the condition of LDM for audio generation.

\subsubsection{Onset-driven Local Temporal Feature}\label{onset}
Temporal alignment is a key aspect that distinguishes V2A from T2A tasks. 
SyncFusion \cite{comunita2024syncfusion} addresses this issue by predicting action onsets in videos and using them as conditions for generating audio. 
However, for universal V2A tasks, sound events in videos are more complex and diverse, and many videos lack onset labels, making it difficult to predict onsets of audio events accurately.

To tackle this challenge, we introduce a pretext task that predicts pseudo-labels for onsets in more generic audio, allowing for learning local video features more related to audio events. 
The pseudo-labels are generated from audio using an Onset Detection Algorithm \cite{bock2013maximum,yariv2024diverse}. 
We first adjust feature dimensions and temporal scale to obtain an embedding with the same temporal length as the audio latent representation $z$. 
Next, we apply an expanding context window technique to capture diverse local features and obtain a new feature $e_{lv}$. 
Finally, we use a linear layer to get logits and calculate the Binary Cross Entropy (BCE) loss with the onset pseudo-labels extracted from the audio.
In this way, by predicting the pseudo-labels of audio onset from video, we acquire local video features $e_{lv}$ that are more temporally aligned with the audio.
\begin{equation}
    \mathcal{L}_{onset} = -\frac{1}{T'}\sum_{i=1}^{T'}{[y_{a}^i log(\hat{y}_{v}^i)+(1-y_{a}^i)(log(1-\hat{y}_{v}^i)]},
\end{equation}
where $y_{a}$ denotes the pseudo-label and $\hat{y}_{v}$ represents the prediction.

\subsubsection{Attentive Pooling Global Semantic Feature}\label{vfe}
In addition to audio-related temporal features, video features contain semantic features, which can be considered global information. Extracting global video features as a condition for audio generation can help enhance the semantic consistency of the generated audio.

First, we use a trainable attentive pooling layer \cite{schwartz2019factor, yariv2024diverse} to aggregate the video features $e_v$. Thus,
\begin{equation}
    \Tilde{e}_v^{atten}=\sum_{u=1}^L p(u)\Tilde{e}_v^{(u)},
\end{equation}
where $p(u)\ge 0\ \forall u$ is a probability distribution, and $\Tilde{e}_v^{(u)}$ represents the $u$'th frame of the video features:
\begin{equation}
    p(u) \propto exp(\alpha_l\theta_l(u)+\alpha_c\theta_c(u)).
\end{equation}
The local potential is $\theta_l(u) = v_l^\mathrm{T} relu(V_l \Tilde{e_v}^{(u)})$, and the cross potential between the audio components is:
\begin{equation}
    \theta_c(u) = \sum_{i=1}^L((\frac{W_1 \Tilde{e}_v^{(u)}}{\Vert W_1 \Tilde{e}_v^{(u)} \Vert})^\mathrm{T}(\frac{W_2 \Tilde{e}_v^{(i)}}{\Vert W_2 \Tilde{e}_v^{(i)} \Vert})),
\end{equation}
where $V_l, W_1, W_2$ are trainable parameters, $v_l$ scores the video component, $\alpha_l, \alpha_c$ calibrates the local and cross potentials. The attention mechanism enables $\Tilde{e_v}^{atten}$ to learn the significance of the video components. Followed by a Conv1D layer, we can get $K$ global video feature $e_{gv}$ by $\Tilde{e_v}^{atten}$ ($K = 4$ by default).

\subsection{T2A-Enhanced Cross-Modal Latent Diffusion Model}\label{base_model}
\subsubsection{T2A Prior Knowledge Initialization}
In recent years, T2A research has experienced rapid development. High-quality audio datasets with text descriptions and superior open-source T2A models are available. Given that V2A, as an audio generation task, necessitates a similar audio distribution with T2A, we aim to capitalize on this aspect fully. Consequently, we adopt the LDM with a mel-spectrogram VAE as the foundational model for STA-V2A. This enables STA-V2A to be initialized with a pre-trained T2A model, thereby retaining its robust audio generation capabilities, reducing the complexity of model training, and enhancing the quality of generated audio.

\subsubsection{Cross-modal Guidance LDM}\label{ldm}
The semantic information of audio potentially comes from both text and video, and there may be differences between the two modalities. 
Therefore, we integrate text and global video information as the guidance types inspired by Uni-ControlNet \cite{zhao2024uni}.
LDM learns the reverse process of a fixed-length Markov Chain of diffusion with condition $c$.
The condition $c$ is concatenated by the pre-trained text embedding $e_{text}$ and the global semantic video features $e_{gv}$, which serves as the key and value of cross attention.
\begin{equation}
\resizebox{0.9\linewidth}{!}{$
    c = [e_{text};e_{gv}] = [e_{text}^1, e_{text}^2, ..., e_{text}^{K_0}; e_{gv}^{1}, \dots, e_{gv}^{K}],
$}
\end{equation}
where [;] represents the concatenation operation and $K_0$ is the length of the original text embedding.
The forward process gradually adds Gaussian noise $\mathcal{N}(0, 1)$ to $z_0$.
The reverse process uses the following loss to denoise and reconstruct $z_0$ through a noise estimation network ($\hat{\epsilon}_\theta$), conditioned on $c$. 
The diffusion loss is as follows:
\begin{equation}
\resizebox{0.9\linewidth}{!}{$
    \mathcal{L}_{DM} = \mathbb{E}_{z_0, \epsilon \sim \mathcal{N}(0,I), t\sim Uniform(1,T)}{\parallel \epsilon-\hat{\epsilon}_\theta(z_t,t,[e_{text};e_{gv}])\parallel_2^2}.
$}
\end{equation}
The noise estimation network ($\hat{\epsilon}_\theta$) is parameterized with U-Net with a cross-attention component to include the condition $c$.
We employ a classifier-free guidance\cite{ho2022classifier} of $e_{text}$ in the reverse process:
\begin{equation}
\resizebox{0.9\linewidth}{!}{$
    \hat{\epsilon}_\theta^{t}(z_t,t,c) = w \cdot \epsilon_\theta^{t}(z_t,t,[e_{text};e_{gv}]) + (1-w) \cdot \epsilon_\theta^{t}(z_t,t,[\phi;e_{gv}]),
$}
\end{equation}
where $\phi$ is the null text embedding and $w$ denotes the guidance scale.
The temporal information of the audio can only be obtained from the video.
We incorporate local temporal video conditions through ControlNet \cite{zhang2023adding} to improve the temporal consistency of generated audio. Unlike conditions for Text-to-Image \cite{zhang2023adding, xuan2024controlnet, zavadski2023controlnet, zhao2024uni} generation, such as depth maps, there is a considerable gap between audio and video modalities. Therefore, the base diffusion model initialized with a T2A model should be trained together with ControlNet. We add noise latent representation $z$ to the local temporal video features $e_{lv}$ learned through the onset prediction pretext task, serving as the input of ControlNet. 

\subsection{New Evaluation Metric for Audio Temporal Alignment}\label{aa-align}
We found that there is a lack of effective metrics for measuring the temporal alignment between audio and audio. Inspired by Audio-Video Alignment (AV-Align) \cite{yariv2024diverse}, we propose a new objective metric AA-Align for the time alignment evaluation of audio generation.
First, we detect peaks in both generated and ground truth audio, denoting the peak sets as $\mathcal{A}_{gen}$ and $\mathcal{A}_{gt}$, respectively.
Then, we verify whether a generated audio peak appears within $T$s ($T = 0.1$ in our paper) before and after the ground truth audio peak, denoted as $1[p_{gen} \in \mathcal{A}_{gt}]$.
Finally, we normalize by the number of peaks to obtain an alignment score between 0 and 1. 
This metric reflects the temporal consistency between generated and ground truth audio. 
More formally, given $\mathcal{A}_{gen}$ and $\mathcal{A}_{gt}$, the alignment score is defined as follows:
\begin{equation}
    \mbox{AA-Align} = \frac{1}{|\mathcal{A}_{gt} \cup \mathcal{A}_{gen}|}\sum_{p_{gen} \in \mathcal{A}_{gen}}1[p_{gen} \in \mathcal{A}_{gt}].
\end{equation}
We consider it a valid peak if the peak of the generated audio is within the $2T$ window of the ground truth audio. 
The above metric can be interpreted as an Intersection-over-Union metric.
\section{Experiments}
\label{sec:exp}
\subsection{Experiment Setup}
\label{setup}

\noindent \textbf{Dataset and Preprocessing.}
We perform our main V2A generation experiments on a subset of the VGGSound \cite{chen2020vggsound} dataset. 
VGGSound contains over 200k clips for 309 different sound classes extracted from YouTube. 
We follow types like Auto-ACD \cite{sun2023large} to generate captions for VGGSound.
We employ two types of video data filtering strategies. First, we use video captions to filter out videos containing human speech.
Second, we utilize the AV-Align \cite{yariv2024diverse} metric to filter out low-quality videos with audio-visual misalignment because we find that some videos exhibit severe audio-visual asynchrony, such as videos with a static, unchanging image. We fix a bug in the AV-Align algorithm by ignoring the local maxima of optical flow less than 0.1, thus preventing static scenes from being incorrectly calculated as peaks due to very small optical flow.
We calculated the corrected AV-Align scores for all videos and found that aside from the peak with a mean of 0, the scores are very close to a normal distribution with a mean of 0.20.
Therefore, we filtered out videos with AV-Align scores less than 0.2, leaving 53,293 samples.

\noindent \textbf{Implementation Details.}
We first pre-train a T2A model on 50,000 hours of YouTube videos and fine-tune it on VGGSound for 40 epochs with a batch size of 160 to initialize our model. The text encoder is a frozen mt5-large \cite{xue2020mt5} text encoder, while the diffusion model is based on the stable diffusion U-Net architecture with 8 channels and a cross-attention dimension of 1024. For the training of STA-V2A, we utilize the AdamW optimizer with a learning rate of 3e-5 and employ 8 V100 GPUs for training, with a per-GPU batch size of 10 and two gradient accumulation steps. Each model has been trained for 40 epochs, and the results are reported for the checkpoint with the best validation loss. During inference, the denoise steps are set to 200, and the guidance scale of classifier-free guidance is 3.

\noindent \textbf{Metrics.}
We utilize objective and subjective metrics to evaluate the performance of the models over audio quality, semantic consistency, and temporal alignment.

For objective evaluation, we use a set of commonly used metrics: Fréchet distance (FD), Fréchet Audio Distance (FAD), KL divergence (KL), Inception Score (IS), Prompting Audio-Language Models (PAM) \cite{deshmukh2024pam}, contrastive language-audio pretraining (CLAP) \cite{wu2023large}, AV-Align (AV) \cite{yariv2024diverse}, and AA-Align (AA).
IS and PAM are effective in evaluating the quality of audio. 
FD and FAD are used to measure the similarity between two audio samples.
KL calculates the divergence between the distributions of two audio. 
CLAP score calculates the degree of match between the generated audio and the video-text description.
AV-Align and AA-Align calculate the temporal alignment between the generated audio and the input video or ground truth audio, respectively.

For subjective evaluation, we conduct crowd-sourced human evaluations, inviting 6 professional annotators to rate the overall quality (OQ), audio quality (AQ), video-audio semantic alignment (SA), and video-audio temporal alignment (TA) of the generated audio, with scores ranging from 1 to 100. For each method, we randomly selected 20 video-audio pairs, all cropped to the same duration. % (4 seconds).
We report OQ, AQ, SA, and TA with 95\% confidence intervals (CI).

\begin{table}[t]
    \caption{Objective metrics of our proposed model and baselines.}
    \begin{center}
    \setlength{\tabcolsep}{0.9mm}{
    \scalebox{0.85}{
    \begin{tabular}{lccccccccc}
        \toprule
        Model & Dur & FD \textcolor{green}{$\downarrow$} & FAD \textcolor{green}{$\downarrow$} & KL \textcolor{green}{$\downarrow$}& IS \textcolor{red}{$\uparrow$} & PAM \textcolor{red}{$\uparrow$} & CLAP \textcolor{red}{$\uparrow$} & AV \textcolor{red}{$\uparrow$} & AA \textcolor{red}{$\uparrow$} \\
        \midrule
        GT                                  & 10s &   -   &   -  &  -   &  -    & 0.319 & 0.485 & 0.300 & 1.000 \\
        \midrule
        Im2Wav                              & & 21.34 & 8.70 & 4.68 & 7.23 & 0.185 & 0.307 & 0.281 & 0.729 \\ 
        % Diff-Foley                          &  & 27.87 & 9.81 & 6.12 & 7.13 & 0.155 & 0.271 & 0.247 & 0.600 \\ 
        \rowcolor{gray!40} 
        % \cellcolor[HTML]{96FFFB}
        STA-V2A (Ours)                       & \multirow{-2}{*}{4s} & \textbf{10.83} & \textbf{2.16} & \textbf{2.61} & \textbf{12.32} & \textbf{0.389}  & \textbf{0.448}    & \textbf{0.297}    & \textbf{0.740} \\ \cline{2-10} 
        Diff-Foley                          &  & 36.98 & 9.73 & 6.76 & 8.19 & 0.205 & 0.278 & 0.215 & 0.517 \\
        \rowcolor{gray!40} 
        % \rowcolor[HTML]{96FFFB}
        STA-V2A (Ours)                       & \multirow{-2}{*}{8s} & \textbf{12.78} & \textbf{1.91} & \textbf{2.59} & \textbf{13.90} & \textbf{0.428}  & \textbf{0.469}    & \textbf{0.289}    & \textbf{0.704} \\ \cline{2-10}
        Seeing\&Hearing                     & \multirow{6}{*}{10s} & 32.92 & 7.32 & 2.62 & 5.83 & - & - & - & - \\
        T2AV                                &  & 33.29 & 4.05 & \textbf{2.12} & 8.02 & - & - & - & - \\
        FoleyCrafter w/o T  & & 27.00 & 4.44 & 4.57 & 9.44 & 0.307 & 0.179 & 0.239 & 0.559 \\
        % FoleyCrafter (w T) & & 29.45 & 5.85 & 10.01& 7.25 & \textbf{0.358} & 0.362 & 0.236 & 0.566 \\
        FoleyCrafter w. T & & 28.13 & 3.45 & 3.56 & 10.70 & \textbf{0.388} & 0.473 & 0.242 & 0.569 \\
        VTA-LDM &  & 25.64 & 2.44 & 3.41 & 10.07 & 0.241 & 0.412 & 0.247 & 0.601 \\
        \rowcolor{gray!40} 
        % \cellcolor[HTML]{96FFFB}
        STA-V2A (Ours)                       &  & \textbf{21.24} & \textbf{1.83} & 2.50 & \textbf{13.45} & 0.276  & \textbf{0.507} & \textbf{0.279} & \textbf{0.687}\\
        \bottomrule
    \end{tabular}
    }
    }
\label{tbl: objective}
\end{center}
\end{table}

\begin{table}[t]
    \centering
    \caption{Subjective metrics of our proposed model and baselines.}
    \scalebox{0.9}{
    \begin{tabular}{l|cccc}
        \toprule
        Model & AQ \textcolor{red}{$\uparrow$} & SA \textcolor{red}{$\uparrow$} & TA \textcolor{red}{$\uparrow$} & OQ \textcolor{red}{$\uparrow$} \\
        \midrule
        GT                                  & $93.33_{\pm 0.87}$ & $94.92_{\pm 0.77}$ & $94.07_{\pm 0.73}$ & $94.38_{\pm 0.77}$ \\
        \hline
        Im2Wav                              & $82.14_{\pm 1.29}$ & $87.96_{\pm 1.25}$ & $83.73_{\pm 1.71}$ & $84.94_{\pm 1.25}$ \\ 
        Diff-Foley                          & $80.21_{\pm 2.34}$ & $80.93_{\pm 4.00}$ & $79.85_{\pm 3.53}$ & $80.53_{\pm 3.17}$ \\ 
        FoleyCrafter w/o T & $85.78_{\pm 1.01}$ & $83.85_{\pm 2.95}$ & $80.39_{\pm 3.10}$ & $83.09_{\pm 2.37}$ \\
        % FoleyCrafter (w T) & $76.50_{\pm 2.44}$ & $55.77_{\pm 5.27}$ & $53.53_{\pm 5.22}$ & $60.36_{\pm 4.43}$ \\
        FoleyCrafter w. T & $87.71_{\pm 1.13}$ & $89.24_{\pm 1.70}$ & $86.17_{\pm 1.81}$ & $87.30_{\pm 1.62}$ \\
        VTA-LDM & $86.67_{\pm 1.31}$ & $89.15_{\pm 1.62}$ & $85.91_{\pm 2.02}$ & $86.69_{\pm 1.65}$ \\
        \rowcolor{gray!40} STA-V2A (Ours)    & \textbf{90.90$_{\pm 1.46}$} & \textbf{93.04$_{\pm 1.36}$}& \textbf{91.05$_{\pm 1.41}$} &\textbf{92.00$_{\pm 1.03}$} \\
        \bottomrule
    \end{tabular}
    }
    \label{tbl:sbj}
    % }
\end{table}

\noindent \textbf{Baseline Models.}
In our study, we examine six advanced V2A models: Im2Wav \cite{sheffer2023hear}, Diff-Foley \cite{luo2024diff}, Seeing\&Hearing \cite{xing2024seeing}, T2AV \cite{mo2024text}, VTA-LDM\cite{xu2024video}, and FoleyCrafter\cite{zhang2024foleycrafter}.
For IM2WAV, Diff-Foley, VTA-LDM, and FoleyCrafter, we use the pre-trained model to evaluate our test set as a baseline. 
For FoleyCrafter, we evaluated both with and without (FoleyCrafter w. or w/o T) text as a condition.
For Seeing\&Hearing and T2AV, we adopt the score reported by their original papers as the codes are not publicly released.

\subsection{Experiment Results and Analysis}
\label{experiment_results}
\subsubsection{Main Results}
\label{results}
We report the objective and subjective metrics results of different models in Table \ref{tbl: objective} and Table \ref{tbl:sbj}. We compare our proposed method STA-V2A with baselines.
Given that Im2wav and Diff-Foley can only generate 4-second and 8-second audio respectively, we clipped the audio generated by our model for a fair comparison.
Table \ref{tbl: objective} demonstrates that our model surpasses Im2wav, Diff-Foley, VTA-LDM, and FoleyCrafter in all objective metrics, except for the PAM score below FoleyCrafter.
Compared to the results reported in the Seeing\&Hearing and T2AV papers, our model is far superior to them in all metrics except for KL, which is slightly lower than T2AV. 
The subjective metrics in Table \ref{tbl:sbj} indicate that our model outperforms the baseline models in all four subjective metrics.

\subsubsection{Ablation Study}\label{ablation}

\noindent \textbf{Video Feature.}
We report the performance of the original T2A model used to initialize U-Net. 
\texttt{Pretrained-T2A} refers to the model pre-trained on YouTube data. 
\texttt{FineTuned-T2A} represents the model fine-tuned on VGGSound after pre-training. 
\texttt{CN} is the model that uses \texttt{FineTuned-T2A} for initialization and introduces the video features encoded by CAVP as control conditions through ControlNet. 
Due to the lack of video information, the audio generated by the T2A model has poor alignment with the video on the time scale, so AV-Align and AA-Align scores are low.  
All objective metrics have shown improvement by incorporating pre-trained video CAVP features as a condition to control audio generation.

\noindent \textbf{Data Filter.} We examine the effect of data filtering by AV-Align scores. 
\texttt{CN w/o Filter} represents the results of training on unfiltered data. 
The experimental results show that compared to \texttt{CN}, training on unfiltered data results in a slight increase in FD, FAD, and KL and a slight decrease in IS and PAM.
The AV-Align decreases by 0.042, and AA-Align decreases by 0.096 when trained on unfiltered data. 
This demonstrates that data filtering improves temporal alignment.
% between the generated audio and the original video.

\noindent \textbf{U-Net Frozen.} 
Due to the gap in video and audio modalities, U-Net copied from the pre-trained model needs to be trained along with the adapter.
We compare the results with and without freezing the U-Net of the original diffusion model. 
The experimental results show that after freezing the U-Net (\texttt{CN Frozen}), IS, PAM, AV-Align, and AA-Align decrease obviously.
Therefore, we did not freeze the parameters of the original diffusion model and trained them together with the adapter.

\noindent \textbf{Onset-driven Local Temporal Feature.} After introducing the onset prediction pretext task to get local temporal features of video, compared to \texttt{CN}, \texttt{CN+Onset} has an increased AV-Align by 0.006 and an increased AA-Align by 0.023. 
This indicates that the onset prediction pretext task can enrich the extracted video features with greater temporal relevance, thereby improving the temporal relevance of the generated audio concerning the video.

\noindent \textbf{Attentive Pooling Global Semantic Feature.} The introduction of the onset prediction pretext task causes the CAVP features to lose some video semantic features. 
The addition of the global video feature extracted in Section \ref{vfe} offsets this, resulting in a decreased FAD by 0.12, an increased IS by 0.22, and an increased PAM by 0.021 with the cost of a slight decrease in other metrics (\texttt{CN+Onset+GVF}).
% For subjective metrics, Table \ref{tbl:abs_sbj} shows that SA, TA, and QQ have increased except for a slight decrease in AQ after adding the video feature enhancement module.
Overall, \texttt{CN+Onset+GVF} (\texttt{STA-V2A}) achieves a good balance between audio quality, semantic consistency, and temporal alignment, yielding the best performance. 

\begin{table}[t]
    \caption{Ablation study for objective metrics of our models. CN+Onset+GVF represents the proposed STA-V2A.}
    \begin{center}
    \setlength{\tabcolsep}{0.9mm}{
    \scalebox{0.92}{
    \begin{tabular}{lccccccccc}
        \toprule
        Model & FD \textcolor{green}{$\downarrow$} & FAD \textcolor{green}{$\downarrow$} & KL \textcolor{green}{$\downarrow$}& IS \textcolor{red}{$\uparrow$} & PAM \textcolor{red}{$\uparrow$} & CLAP \textcolor{red}{$\uparrow$} & AV \textcolor{red}{$\uparrow$} & AA \textcolor{red}{$\uparrow$} \\
        \midrule
        GT                                  &   -   &   -  &  -   &  -    & 0.319 & 0.485 & 0.300 & 1.000 \\
        \midrule
        Pretrained-T2A                      & 27.25 & 3.53 & 3.64 & 10.32 & 0.258 & 0.470 & 0.206 & 0.498 \\ 
        FineTuned-T2A                       & 27.24 & 3.84 & 3.57 & 10.44 & 0.260 & 0.472 & 0.208 & 0.502 \\ 
        CN w/o Filter                       & 19.65 & 2.00 & 2.53 & 13.15 & 0.260 & 0.508 & 0.235 & 0.573 \\ 
        CN Frozen                           & 20.35 & 1.90 & 2.50 & 12.26 & 0.245 & 0.510 & 0.269 & 0.649 \\ 
        CN                                  & \textbf{19.58} & 2.23 & 2.51 & 13.38 & 0.273 & 0.506 & 0.277 & 0.669\\ 
        CN+Onset                            & 20.02 & 1.95 & \textbf{2.37} & 13.23 & 0.255 & \textbf{0.512} & \textbf{0.283} & \textbf{0.692} \\  
        \rowcolor{gray!40} 
        % \cellcolor[HTML]{96FFFB}
        CN+Onset+GVF     & 21.24 & \textbf{1.83} & 2.50 & \textbf{13.45} & \textbf{0.276}  & 0.507 & 0.279 & 0.687\\
        \bottomrule
    \end{tabular}
    }
    }
\label{tbl:abs}
\end{center}
\end{table}

\section{Conclusion}
\label{sec:conclusion}
We introduce STA-V2A, an approach designed to generate semantically and temporally aligned audio for video. STA-V2A leverages both text and video features as conditions. For video features, we propose an onset prediction pretext task and a trainable attentive pooling module to extract local temporal and global semantic features, effectively reducing the interference from redundant information within the video features. Besides, we propose the T2A-Enhanced Cross-Modal LDM, which simultaneously improves the quality, semantic alignment, and temporal alignment of the generated audio by employing T2A initialization and cross-modal conditioning. Furthermore, the proposed AA-Align evaluation metric is an effective metric for the temporal alignment of audio generation. Finally, extensive experiments demonstrate that STA-VTA has achieved significant advancements in semantic and temporal alignment, and the ablation analysis validates the effectiveness of the proposed modules.

\vfill\pagebreak

\bibliographystyle{IEEEtran}
% \bibliography{IEEEabrv,refs}
\bibliography{refs}

\end{document}